\documentclass[conference,a4paper]{APSIPA2025}
\usepackage{amsmath}
\usepackage{graphicx}
\usepackage{multirow}
\usepackage{threeparttable}
\usepackage[backend=biber,style=ieee,]{biblatex}
\usepackage{booktabs}
\usepackage[normalem]{ulem}
\usepackage{amssymb}
\useunder{\uline}{\ul}{}
\usepackage{geometry}
\usepackage[T1]{fontenc}
\usepackage{etoolbox}
\AtBeginBibliography{\fontsize{9}{9.9}\selectfont}
\usepackage{fancyhdr}

\fancypagestyle{firststyle}{
  \fancyhf{}
  \fancyhead[C]{2025 Asia Pacific Signal and Information Processing Association Annual Summit and Conference (APSIPA ASC)}
}

\begin{document}

\title{Interpolating Speaker Identities in Embedding Space for Data Expansion}

\author{
\authorblockN{
{\fontsize{10.9pt}{11pt}\selectfont{Tianchi Liu\authorrefmark{1}}},
{\fontsize{10.9pt}{11pt}\selectfont{Ruijie Tao\authorrefmark{1}}},
{\fontsize{10.9pt}{11pt}\selectfont{Qiongqiong Wang\authorrefmark{2}}},
{\fontsize{10.9pt}{11pt}\selectfont{Yidi Jiang\authorrefmark{1}}},
{\fontsize{10.9pt}{11pt}\selectfont{{Hardik B. Sailor}\authorrefmark{2}}},
{\fontsize{10.9pt}{11pt}\selectfont{Ke Zhang\authorrefmark{3}}},
{\fontsize{10.9pt}{11pt}\selectfont{Jingru Lin\authorrefmark{1}}},
{\fontsize{10.9pt}{11pt}\selectfont{Haizhou Li\authorrefmark{3}\authorrefmark{1}}}
}

% \authorblockA{
% \authorrefmark{1}
% LIGHTSPEED, Singapore 
% }

\authorblockA{
\authorrefmark{1}
Department of Electrical and Computer Engineering, National University of Singapore, Singapore 
% \\
% \small E-mail: ruijie.tao@u.nus.edu, yidi\underline{\enskip}jiang@u.nus.edu
}

\authorblockA{
\authorrefmark{2}
Institute for Infocomm Research (I$^{2}$R), Agency for Science, Technology and Research (A$^\star$STAR), Singapore 
% \\
% \small E-mail: wang\underline{\enskip}qiongqiong@i2r.a-star.edu.sg, sailor\underline{\enskip}hardik\underline{\enskip}bhupendra@i2r.a-star.edu.sg
}

\authorblockA{
\authorrefmark{3}
SRIBD, School of Artificial Intelligence, The Chinese University of Hong Kong, Shenzhen, China
\\
\small E-mail: tianchi\_liu@u.nus.edu
% \\
% \small E-mail: kezhang@cuhk.edu.cn, haizhouli@cuhk.edu.cn
}
}

\maketitle
\thispagestyle{firststyle}
\pagestyle{fancy}

\begin{abstract}
The success of deep learning-based speaker verification systems is largely attributed to access to large-scale and diverse speaker identity data. However, collecting data from more identities is expensive, challenging, and often limited by privacy concerns. To address this limitation, we propose INSIDE (Interpolating Speaker Identities in Embedding Space), a novel data expansion method that synthesizes new speaker identities by interpolating between existing speaker embeddings. Specifically, we select pairs of nearby speaker embeddings from a pretrained speaker embedding space and compute intermediate embeddings using spherical linear interpolation. These interpolated embeddings are then fed to a text-to-speech system to generate corresponding speech waveforms. The resulting data is combined with the original dataset to train downstream models. Experiments show that models trained with INSIDE-expanded data outperform those trained only on real data, achieving 3.06\% to 5.24\% relative improvements. While INSIDE is primarily designed for speaker verification, we also validate its effectiveness on gender classification, where it yields a 13.44\% relative improvement. Moreover, INSIDE is compatible with other augmentation techniques and can serve as a flexible, scalable addition to existing training pipelines.

\end{abstract}

\section{Introduction}
Speaker verification (SV) is the task of determining whether a given speech segment belongs to a claimed speaker~\cite{KINNUNEN201012}. Recent progress in deep learning has significantly advanced this task, resulting in remarkable performance gains~\cite{yakovlev24_interspeech, Golden_Gemini, 10887949, 10889058, 9681187}. The availability of large-scale, diverse, and labeled datasets has played an important role in enabling models to learn robust and discriminative speaker representations~\cite{Voxblink, 3dspeaker, 10760244}. However, collecting and annotating such datasets is often resource-intensive, time-consuming, and fraught with privacy concerns~\cite{Synvox2}.

Previous studies have widely employed techniques such as additive noise, reverberation, and speed perturbation to improve robustness in speaker verification~\cite{X-Vector}. These techniques operate at the acoustic level and do not generate new speaker identities. An exception is speed perturbation~\cite{SpeedPerturbation}, which has been argued by Yamamoto et al.~\cite{SpeedPerturbation_SPK} to create new speaker identities. However, this method does not fundamentally modify speaker-specific characteristics, thereby limiting the diversity of speaker identities in the training data.

To overcome these limitations, we propose a more flexible and effective approach that generates synthetic speaker identities directly in the speaker embedding space, rather than relying on superficial acoustic transformations. Speaker embeddings, which capture speaker-specific characteristics in a learned latent space, have become a core component in modern speaker recognition systems. Importantly, the geometric structure of this space allows for meaningful interpolation between different speaker embeddings. Inspired by this property, we introduce \textbf{INSIDE} (\textbf{In}terpolating \textbf{S}peaker \textbf{Id}entities in \textbf{E}mbedding Space), a data expansion method that creates new speaker identities by interpolating between embeddings of real speakers. These interpolated embeddings are then used to condition a text-to-speech (TTS) model, enabling the generation of synthetic speech samples that represent new speaker identities. The resulting data expands the diversity of training speakers, enhancing model robustness and generalization.
The INSIDE framework offers several advantages:

\begin{itemize}
    \item The proposed INSIDE is a scalable and privacy-friendly data expansion method that generates diverse speaker identities without requiring additional data collection.
    \item As the synthetic speakers are created through interpolation in the embedding space, the resulting samples preserve the semantic structure of speaker characteristics, which leads to more stable and effective model training.
    \item The method is controllable, allowing flexible adjustment of gender ratio, language, content, and the number of identities to match specific augmentation needs.
    \item While our primary focus is speaker verification, we also validate INSIDE on gender classification. The framework shows potential for broader speech-related tasks.
\end{itemize}

\begin{figure*}[t]
\centering
\includegraphics[width=0.86\textwidth]{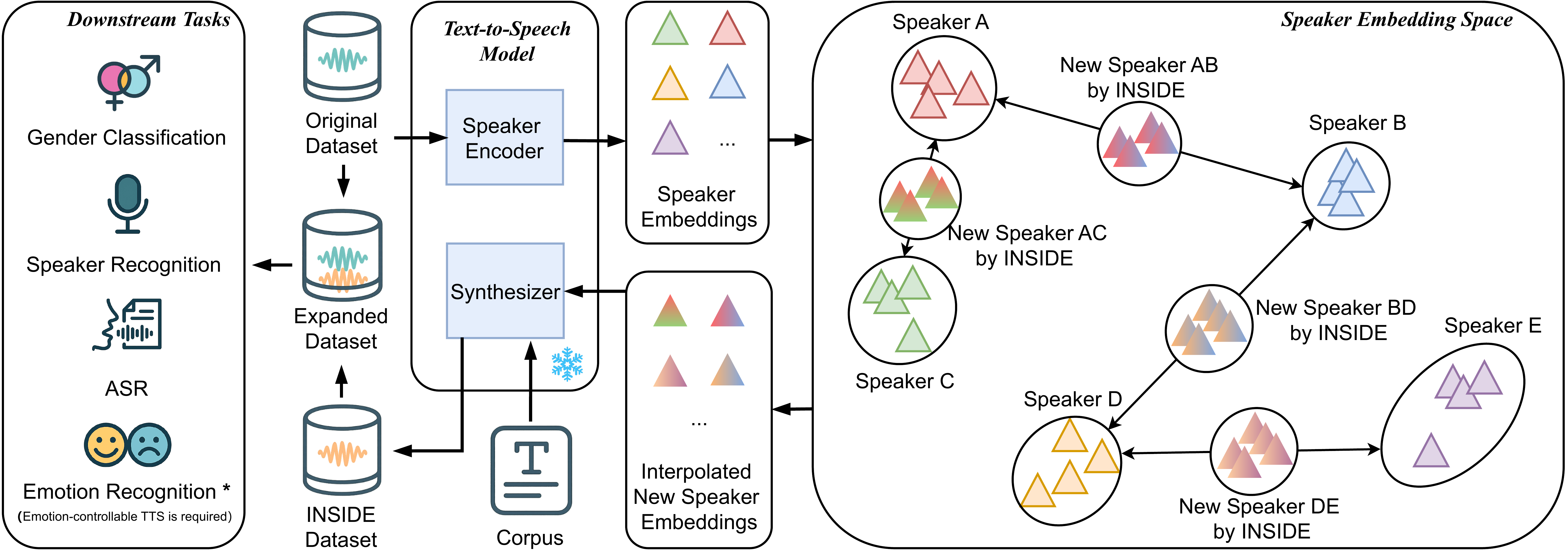}
\vspace{-0.05 in}
\caption{Overview of the INSIDE data expansion pipeline. The snowflake icon denotes that the TTS model is pretrained and remains frozen.}
\label{fig_INSIDE}
\vspace{-0.15 in}
\end{figure*}

\section{Related Work}
\subsection{Data Augmentation in Speaker Verification}

Data augmentation plays a key role in improving the robustness of speaker verification systems. Most existing methods focus on acoustic-level transformations, such as adding noise and reverberation~\cite{X-Vector, 9167416}, or using time-domain techniques like speed perturbation~\cite{SpeedPerturbation}. Speed perturbation alters the speaking rate of recordings, typically by applying speed factors. This method introduces variability and is sometimes used to simulate new speaker identities~\cite{SpeedPerturbation_SPK}.
However, these approaches have clear limitations. Since speed perturbation and other acoustic-level methods only modify low-level signal characteristics, the generated utterances remain tightly coupled with the original speakers and phonetic content. They neither create truly new identities nor introduce sufficient variability in linguistic or phonetic expression. As a result, such methods struggle to capture intra-speaker variation and provide limited benefits when scaling to unseen speakers.

To address these limitations, we propose INSIDE, a framework that augments data directly in the embedding space. It generates new speaker identities by interpolating between real embeddings. This enables flexible, identity-level expansion with independent control over linguistic content, improving both speaker discriminability and intra-speaker variability.

\subsection{Synthetic Speech Dataset}
The advancement of speaker recognition systems relies on large-scale and diverse datasets. However, high collection costs and privacy concerns have led to the usage of synthetic datasets. These datasets aim to supplement or replace real-world data~\cite{Synvox2}, offering benefits such as improved privacy, scalability, and the ability to simulate various conditions.

Synthetic audio has been widely used for data augmentation in many speech tasks~\cite{10889137}. For example, Libri2Vox~\cite{Libri2Vox} supports target speaker extraction, and SynthASR~\cite{fazel21_interspeech} is used in automatic speech recognition (ASR) where annotated data is limited or hard to obtain.
In SV, synthetic data has also been explored. Tao et al.~\cite{tao2024voice} use synthetic data to enhance defective datasets. SynVox2~\cite{Synvox2} addresses privacy risks in datasets like VoxCeleb2 by generating anonymized yet useful synthetic alternatives. SynAug~\cite{SynAug} augments text-dependent speaker verification by synthesizing fixed-transcript utterances conditioned on embeddings of real speakers. While effective, this method still relies on existing speaker identities and is therefore not scalable beyond the enrolled speaker set, limiting identity diversity and raising ongoing privacy concerns.

These constraints motivate the design of INSIDE, an identity-level data expansion framework that generates virtual speaker identities by interpolating between embeddings of real speakers in latent space. This enables scalable and diverse data generation without collecting new identity data, improving generalization while reducing privacy risks.

\section{Method}
\subsection{Overview of the INSIDE}
Diverse speaker identities are essential for speaker verification, while collecting large labeled datasets is expensive and raises privacy concerns~\cite{RecXi, 9747162}.
Since speaker embeddings typically lie in a structured latent space where real speakers are unevenly distributed and certain regions are sparsely populated~\cite{9296778}, INSIDE leverages this property to generate intermediate embeddings between similar speakers, effectively filling underrepresented areas. This results in improved coverage of the embedding space and exposes the model to richer speaker variation, improving generalization. These new identities also smooth decision boundaries, helping reduce overfitting. As a result, INSIDE enhances both data diversity and distribution, leading to more reliable speaker verification.

\subsection{Interpolating Speaker Identities in Embedding Space}
\label{subsec_INSIDE_ori}
The proposed INSIDE pipeline is illustrated in Fig.~\ref{fig_INSIDE}. We first use a pretrained speaker encoder from a controllable TTS model to extract averaged speaker embeddings $\mathbf{e}_i \in \mathbb{R}^N$ from a labeled dataset, where $N$ is the embedding dimension.

To ensure natural interpolation, we group speakers by gender to avoid blending across major acoustic boundaries and preserve coherence in the generated identities. Let $\mathcal{E}$ be the set of embeddings with the same gender as $\mathbf{e}_i$, from which identity pairs $(\mathbf{e}_i, \mathbf{e}_j)$ are sampled for the interpolation process.

Given two source embeddings, several interpolation strategies can be considered, such as linear interpolation and spherical linear interpolation (SLERP)~\cite{SLERP}. We adopt SLERP because it better fits the hyperspherical geometry of embedding spaces and preserves unit norm, ensuring that interpolated embeddings lie on the unit hypersphere, which aligns with the cosine similarity metric commonly used in speaker verification~\cite{10096626}. The SLERP operation is defined as:
\begin{equation}
\theta = \arccos \left( \frac{\mathbf{e}_i^\top \mathbf{e}_j}{|\mathbf{e}_i| |\mathbf{e}_j|} \right),
\end{equation}
\begin{equation}
\mathbf{e}_{ij} = \frac{\sin((1 - \alpha)\theta)}{\sin(\theta)} \cdot \mathbf{e}_i + \frac{\sin(\alpha \theta)}{\sin(\theta)} \cdot \mathbf{e}_j,\quad \alpha \in [0,1].
\end{equation}

The coefficient $\alpha$ balances the two source speakers, enabling smooth transitions and avoiding artifacts from naive averaging in angular-sensitive spaces.

As the interpolated embeddings $\mathbf{e}_{ij}$ lie in the same embedding space as the original ones, they can be fed directly into the TTS synthesizer without introducing domain mismatch, as illustrated in Fig.~\ref{fig_INSIDE}.
A set of texts with appropriate lengths is extracted from a large-scale corpus to serve as the linguistic content. For each synthetic speaker, a random subset of these texts is selected, and corresponding speech waveforms are synthesized. This process transforms abstract latent vectors into realistic speech data while preserving the identity attributes encoded in the embedding. The resulting utterances constitute the INSIDE dataset, which can be combined with real data to train downstream models.

\subsection{Optimized Pair Selection via Nearest-Neighbor Traversal}
\label{subsec_INSIDE_opt}

\begin{figure}[ht]
\centering
\vspace{-0.1 in}
\includegraphics[width=0.92\columnwidth]{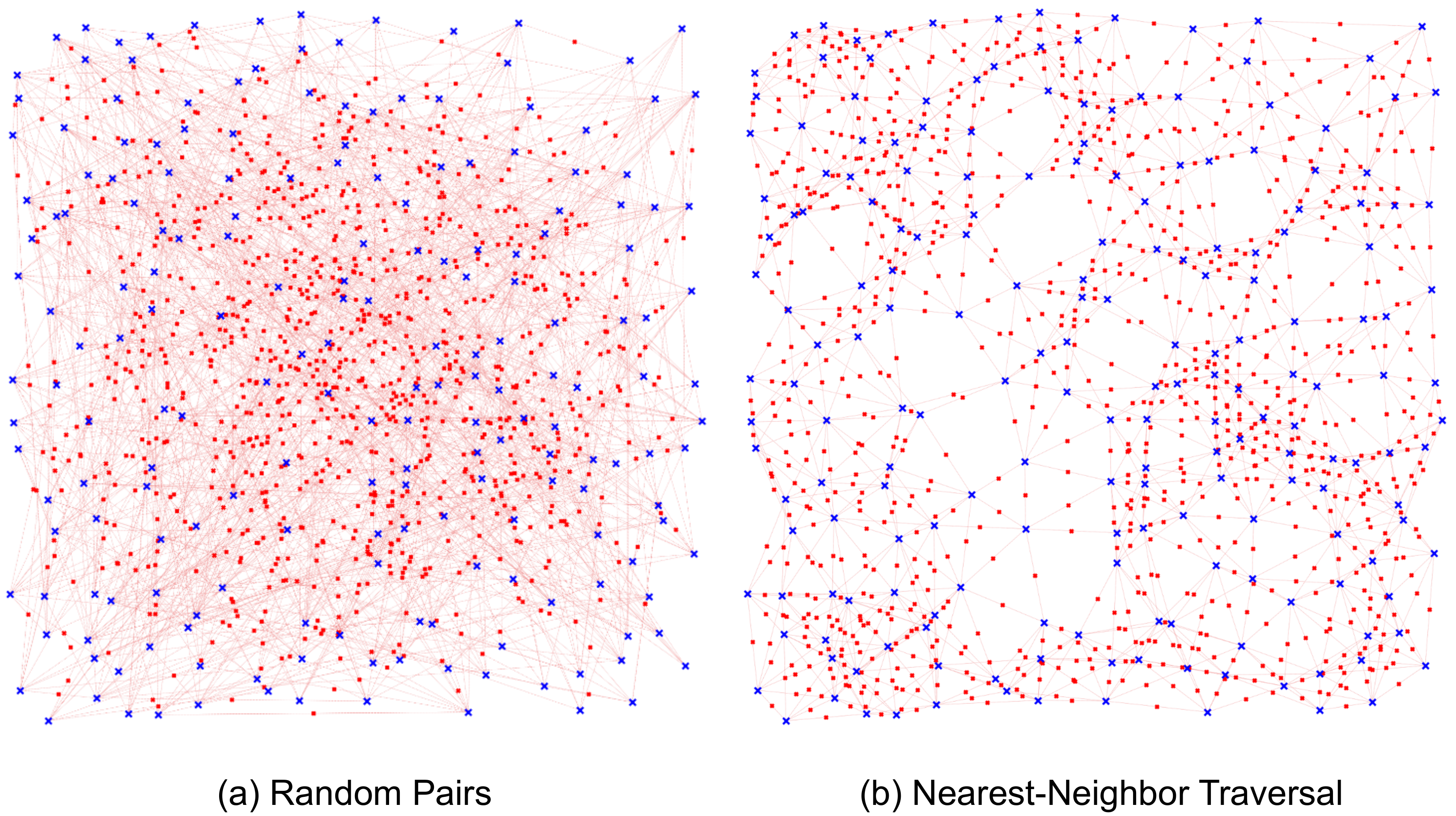}
\vspace{-0.1 in}
\caption{Visualization of identity interpolation in the speaker embedding space. Blue points represent real speaker embeddings. Red lines connect selected pairs of real speakers. Red points denote synthetic speaker identities generated via interpolation between each pair.}
\label{fig_pair}
\vspace{-0.1 in}
\end{figure}

As shown in Fig.~\ref{fig_pair} (a), while random pairing is effective, it results in uneven coverage of the embedding space. Interpolated embeddings $\mathbf{e}_{ij}$ tend to cluster in specific regions of the embedding space, leaving peripheral areas underrepresented.

To mitigate over-sampling in dense central regions and improve overall coverage, we introduce a layered nearest-neighbor interpolation strategy inspired by~\cite{tao2024voice}. Specifically, for each embedding $\mathbf{e}_i$, we compute cosine distances to all other embeddings within the same gender group:
\begin{equation}
d_{ij} = 1 - \frac{\mathbf{e}_i^\top \mathbf{e}_j}{|\mathbf{e}_i| |\mathbf{e}_j|}, \quad i \ne j.
\end{equation}

We build identity pairs by gradually expanding the neighborhood. The $n$ nearest neighbors of $i$ are defined as:
\begin{equation}
\mathcal{N}_n(i) = { j \in \mathcal{E} \mid \text{rank}_i(j) \leq n,\ j \ne i }.
\end{equation}

The candidate pair pool ($\mathcal{P}$) at neighborhood level $n$ is defined as:
\begin{equation}
\mathcal{P}_n = {(i, j) \mid j \in \mathcal{N}_n(i)}.
\end{equation}

To avoid duplication due to symmetry (i.e., both $(i,j)$ and $(j,i)$), we apply a uniqueness constraint. Let $\mathcal{S}$ be the set of selected pairs. We define:
\begin{equation}
\mathcal{S} = \bigcup_{n=1}^{n_{\text{max}}} \text{UniquePairs}(\mathcal{P}_n \setminus \mathcal{S}) \quad \text{s.t.} \quad |\mathcal{S}| \geq T,
\end{equation}
where $\setminus$ denotes the set difference. $T$ is the desired number of synthetic identities. We begin with $n=1$, collecting the closest neighbor for each identity. If the number of unique pairs is insufficient, $n$ is incremented, and the next closest neighbors are included, repeating until $|\mathcal{S}| \geq T$. If the final level $n = k$ produces more candidates than required, we randomly sample the remainder from $\mathcal{P}_k \setminus \mathcal{S}$ to meet the target.

This method promotes a more uniform traversal of the embedding space. As shown in Fig.~\ref{fig_pair} (b), optimized pairing yields a distribution that more closely resembles that of real data compared to random sampling, while also filling some under-represented regions of the original dataset. This leads to smoother decision boundaries during training.

\section{Experimental Setup}
\subsection{Train Set}

\begin{table}[h]
\vspace{-0.15 in}
\scriptsize
\centering
\caption{Statistics of Train Set.}
\label{tab_train_statics}
\vspace{-0.1 in}
\setlength{\tabcolsep}{0.85mm}{
\begin{tabular}{llrrrr}
\hline
\toprule
Train Dataset & Set     & \# of male & \# of female & \# of speakers & \# of samples \\ \hline \midrule
VoxCeleb2 & Dev         & 3,682      & 2,312        & 5,994          & 1,092,009     \\
INSIDE    & Synthetic   & 3,682      & 2,312        & 5,994          & 1,092,009     \\
INSIDE    & Nearest-Neighbor    & 3,682      & 2,312        & 5,994          & 1,092,009     \\
INSIDE    & Identity-expanded & 20,000     & 20,000       & 40,000         & 1,000,000    \\
\bottomrule
\hline
\end{tabular}
}
\vspace{-0.15 in}
\end{table}

We follow the common speaker verification protocol by using the VoxCeleb2~\cite{voxceleb2} dataset as the training set for our baseline models and evaluating performance on the VoxCeleb1~\cite{10096626, chen2025pushing, 9747021}. In addition, we construct three synthetic datasets using our proposed INSIDE:
\begin{itemize}
    \item \textbf{Synthetic (Syn)} comprises synthetic speech generated using the method described in Section~\ref{subsec_INSIDE_ori}, where new speaker identities are created between random pairs.
    \item \textbf{Nearest-Neighbor (NN)} improves upon INSIDE-Syn dataset by optimizing the selection of speaker pairs, as introduced in Section~\ref{subsec_INSIDE_opt}.
    \item \textbf{Identity-Expanded (ID-Exp)} follows the same method as NN but greatly increases the number of speaker identities, aiming to investigate whether further expanding identity quantity leads to additional performance gains.
\end{itemize}

\subsection{Test Set}

\begin{table}[ht]
\scriptsize
\centering
\vspace{-0.05 in}
\caption{Statistics of test sets for speaker verification.}
\vspace{-0.1 in}
\label{tab_spk_test_statics}
\begin{tabular}{lrr}
\hline
\toprule
Test Dataset &     \# of identities &    \# of trial pairs          \\ \hline \midrule
VoxCeleb1-Original (Vox1-O)~\cite{voxceleb1} & 40 & 37,611 \\ 
VoxCeleb1-Extended (Vox1-E)~\cite{voxceleb1}  & 1,251        & 579,818               \\ 
VoxCeleb1-Hard (Vox1-H)~\cite{voxceleb1}     & 1,251       & 550,894          \\

\bottomrule
\hline
\end{tabular}
\vspace{-0.15 in}
\end{table}

\begin{table*}[t]
\centering
\caption{Speaker verification results in EER (\%) and minDCF on the VoxCeleb1 dataset. The best result in each group is shown in \textbf{bold}, and the second-best is {\ul underlined}. `Averaged Improvement' shows the average relative gain over the baseline across all test subsets.}
\vspace{-0.1 in}
\label{tab_spk}
\setlength{\tabcolsep}{1.9 mm}{
\begin{tabular}{cccccccccccc}
\hline
\toprule
 &  &  &  &  & \multicolumn{2}{c}{Vox1-O} & \multicolumn{2}{c}{Vox1-E} & \multicolumn{2}{c}{Vox1-H} &  \\
\multirow{-2}{*}{front-end} & \multirow{-2}{*}{Backend} & \multirow{-2}{*}{\begin{tabular}[c]{@{}c@{}}INSIDE \\ Portion  Added\end{tabular}} & \multirow{-2}{*}{\begin{tabular}[c]{@{}c@{}}Front-end \\ Finetune\end{tabular}} & \multirow{-2}{*}{AS-Norm} & EER↓ & minDCF↓ & EER↓ & minDCF↓ & EER↓ & minDCF↓ & \multirow{-2}{*}{\begin{tabular}[c]{@{}c@{}}Averaged\\ Improvement↑\end{tabular}} \\ \hline \midrule
 &  & N/A &  & \checkmark  & {\ul 0.691} & 0.097 & {\color[HTML]{333333} 0.785} & 0.088 & 1.612 & 0.175 & - \\
 &  & \textbf{Syn} &  & \checkmark  & 0.697 & {\ul 0.089} & 0.780 & {\ul 0.086} & 1.634 & 0.172 & 1.82\% \\
 &  & \textbf{NN} &  & \checkmark  & 0.694 & 0.090 & {\ul 0.758} & 0.092 & \textbf{1.532} & \textbf{0.165} & 2.76\% \\
 & \multirow{-4}{*}{\begin{tabular}[c]{@{}c@{}}ECAPA-TDNN\\ (C=512)\end{tabular}} & \textbf{ID-Exp} &  & \checkmark  & \textbf{0.649} & \textbf{0.089} & \textbf{0.747} & \textbf{0.084} & {\ul 1.569} & {\ul 0.166} & 5.24\% \\ \cline{2-12} 
 &  & N/A &  &  & 0.595 & - & 0.719 & - & 1.501 & - & - \\
 &  & \textbf{ID-Exp} &  &  & \textbf{0.545} & 0.068 & \textbf{0.699} & 0.076 & \textbf{1.488} & 0.144 & 4.02\% \\ \cline{3-12} 
 &  & N/A &  & \checkmark  & 0.548 & - & 0.656 & - & 1.355 & - & - \\
 &  & \textbf{ID-Exp} &  & \checkmark  & \textbf{0.518} & 0.072 & \textbf{0.635} & 0.069 & \textbf{1.334} & 0.137 & 3.38\% \\ \cline{3-12} 
 &  & N/A & \checkmark  &  & 0.542 & - & 0.635 & - & 1.355 & - & - \\
 &  & \textbf{ID-Exp} & \checkmark  &  & \textbf{0.489} & 0.058 & \textbf{0.619} & 0.068 & \textbf{1.337} & 0.136 & 4.53\% \\ \cline{3-12} 
 &  & N/A & \checkmark  & \checkmark  & 0.521 & - & 0.594 & - & \textbf{1.237} & - & - \\
\multirow{-12}{*}{\begin{tabular}[c]{@{}c@{}}WavLM\\ -Large\end{tabular}} & \multirow{-8}{*}{\begin{tabular}[c]{@{}c@{}}ECAPA-TDNN\\ Global\\ (C=512)\end{tabular}} & \textbf{ID-Exp} & \checkmark  & \checkmark  & \textbf{0.479} & 0.059 & \textbf{0.583} & 0.063 & 1.247 & 0.124 & 3.06\% \\
\bottomrule
\hline
\end{tabular}
}
\vspace{-0.2 in}
\end{table*}

\begin{table}[t]
\scriptsize
\centering
\vspace{0.1 in}
\caption{Statistics of test sets for gender classification.}
\vspace{-0.1 in}
\label{tab_gender_statics}
\setlength{\tabcolsep}{1.6mm}{
% \scalebox{0.85}{
\begin{tabular}{llrl}
\hline
\toprule
Test Dataset & Set      & \# of samples & Remarks          \\ \hline \midrule
VoxCeleb1~\cite{voxceleb1}              & -            & 153,516       & In-domain        \\ \hline 
\multirow{2}{*}{TIMIT~\cite{TIMIT}} & Train        & 4,620         & \multirow{2}{*}{\begin{tabular}[c]{@{}l@{}}Cross-domain, English\end{tabular}}      \\ 
         & Test         & 1,680         & \\ \hline
\multirow{4}{*}{Private}  & Part 1 Train & 2,264,026     & \multirow{4}{*}{\begin{tabular}[c]{@{}l@{}}Cross-domain,\\ South-east Asia languages\end{tabular}} \\
         & Part 1 Test  & 5,000         & \\
         & Part 2 Train & 2,473,967     & \\
         & Part 2 Test  & 5,000         & \\ \hline
\begin{tabular}[c]{@{}l@{}}Samrómur Children~\cite{samromur_Children}\end{tabular}      & Train        & 129,446       & \begin{tabular}[c]{@{}l@{}}Cross-domain,\\ Icelandic children speech
\end{tabular}          \\
\bottomrule
\hline
\end{tabular}
}
\vspace{-0.2 in}
\end{table}

For the speaker verification task, as shown in Table~\ref{tab_spk_test_statics}, we evaluate our models on the widely used VoxCeleb1~\cite{voxceleb1} dataset, which includes three standard evaluation protocols: original, extended, and hard cases.

In addition to speaker verification, we investigate whether our proposed method can benefit other speech-related tasks. We use gender classification as a case study and evaluate on three public datasets—VoxCeleb1~\cite{voxceleb1}, TIMIT~\cite{TIMIT}, and Samrómur Children~\cite{samromur_Children}—as well as a private dataset. Since gender classification typically achieves high accuracy, we include these diverse datasets to assess our method under varying acoustic conditions and speaker demographics, aiming to validate its generalizability and robustness.

\subsection{Training Configuration}

To construct the INSIDE datasets, we use $\alpha=0.5$ to generate midpoint embeddings, representing a balanced blend of both identities. The speech synthesis model used is YourTTS~\cite{YourTTS}, provided by Coqui\footnote{\url{https://github.com/coqui-ai/TTS}}. Text samples are drawn from the LibriSpeech~\cite{Librispeech} and filtered to 100–250 words to match the average duration of VoxCeleb2 utterances~\cite{voxceleb2}.

All experiments are conducted using the WeSpeaker toolkit~\cite{10096626}, and the same training strategies are applied to models trained with either the original dataset or its combination with INSIDE data. WavLM-Large~\cite{WavLM} is used as the front-end; reverberation and additive noise are applied with a probability of 0.6; all models are optimized using SGD with momentum 0.9 and weight decay of $1 \times 10^{-4}$.

For speaker verification, the training settings differ by back-end architecture.
Models with the ECAPA-TDNN~\cite{ECAPA} back-end are trained for 40 epochs using AAM-softmax with a scale of 32 and a margin that increases from 0 to 0.3 between epochs 5 and 10. The learning rate decays exponentially from 0.1 to $5 \times 10^{-5}$ with a 2-epoch warm-up. Input utterances are cut to 300 frames, and speed perturbation is applied~\cite{SpeedPerturbation,SpeedPerturbation_SPK}.
For ECAPA-TDNN with global context attention (ECAPA-TDNN Global)~\cite{ECAPA}, we follow the WeSpeaker configurations\footnote{\url{https://github.com/wenet-e2e/wespeaker}} and train for 150 epochs. The learning rate decays exponentially from 0.1 to $1 \times 10^{-5}$ with a 6-epoch warm-up. The loss is AAM-softmax with sub-center, and inter top-k penalty~\cite{zhao2021speakin} with a scale of 32 and a margin that increases exponentially from 0 to 0.2, starting at epoch 20 and fixed after epoch 40. Utterances are cut to 150 frames. Speed perturbation is also applied.
When the front-end model is jointly finetuned with the back-end during training, the model is trained for 20 epochs with the learning rate decaying from 0.001 to $2.5 \times 10^{-4}$ and a 1-epoch warm-up. Other settings remain the same as above.

For gender classification, we use a lightweight ECAPA-TDNN with channel size 128 and softmax loss. The front-end remains frozen. Training lasts for 3 epochs, with the learning rate decaying from 0.1 to $5 \times 10^{-5}$ and no warm-up. Utterances are fixed to 200 frames.

\begin{table*}[t]
\centering
\caption{Gender classification results in error rate (\%). The best result in each test track is shown in \textbf{bold}. `Averaged Imp.' indicates the average relative improvement over the baseline across all test subsets.}
\vspace{-0.05 in}
\label{tab_gender}
\setlength{\tabcolsep}{1.15 mm}{
\begin{tabular}{cccrccccccccc}
\hline
\toprule
& & \multirow{2}{*}{\begin{tabular}[c]{@{}c@{}}INSIDE \\ Portion Added\end{tabular}}      & \multirow{2}{*}{\begin{tabular}[c]{@{}c@{}}\# of\\Total IDs\end{tabular}} & \multirow{2}{*}{\begin{tabular}[c]{@{}c@{}}VoxCeleb1\\~\cite{voxceleb1}\end{tabular}} & \multicolumn{2}{c}{TIMIT~\cite{TIMIT}}     & \multicolumn{4}{c}{Private}          & \multirow{2}{*}{\begin{tabular}[c]{@{}l@{}}Samrómur \\ Children~\cite{samromur_Children} ↓\end{tabular}}& \multirow{2}{*}{\begin{tabular}[c]{@{}l@{}}Averaged\\ Imp. ↑\end{tabular}} \\ \cmidrule(r){6-7} \cmidrule(r){8-11}

\multirow{-2}{*}{front-end} & \multirow{-2}{*}{Backend} & & &              & Train↓         & Test↓         & P1 Train↓      & P1 Test↓       & P2 Train↓      & P2 Test↓       &        &        \\ \hline \midrule
&  \multirow{3}{*}{\begin{tabular}[c]{@{}c@{}}ECAPA-TDNN\\ (C=128)\end{tabular}} & N/A            & 5,994     & 1.24         & 0.50          & 0.89          & 2.20          & 2.42          & 2.55          & 2.36          & 37.01  & -            \\ 
& & \textbf{NN} & 11,988              & \textbf{1.13}              & 0.25          & \textbf{0.60} & 2.16          & 2.46          & 2.55          & 2.42          & 34.17  & 12.09\%                \\
\multirow{-3}{*}{\begin{tabular}[c]{@{}c@{}}WavLM\\ -Large\end{tabular}} & & \textbf{ID-Exp} & 45,994              & 1.42         & \textbf{0.17} & 0.65          & \textbf{2.10} & \textbf{2.26} & \textbf{2.49} &  \textbf{2.30} & \textbf{32.18}       & 13.44\% \\
\bottomrule
\hline
\end{tabular}
\vspace{-0.15 in}
}
\end{table*}
\section{Result}
\subsection{Speaker Verification}
We first evaluate the effectiveness of the proposed INSIDE on the SV task, using WavLM-Large as the front-end and ECAPA-TDNN as the back-end. Our baseline achieves strong performance on the VoxCeleb1 test sets, as shown in  Table~\ref{tab_spk}. Augmenting the training data with synthetic identities generated by the INSIDE approach (Section~\ref{subsec_INSIDE_ori}) yields an average relative improvement of 1.82\%.

Further gains are achieved using the nearest-neighbor optimized pairing strategy (Section~\ref{subsec_INSIDE_opt}), referred to as INSIDE-NN, which improves performance by 2.76\%, indicating that the distribution of synthetic identities has a noticeable impact on the quality of the augmented dataset.

Taking advantage of INSIDE’s scalability, we expand the number of synthetic identities to 40,000, creating the ID-Exp dataset. This leads to a substantial average relative improvement of 5.24\%, demonstrating the importance of identity quantity in speaker-related tasks.

Additionally, we assess the robustness and compatibility of INSIDE within the WeSpeaker~\cite{10096626} framework under various settings, including whether front-end finetuning and AS-Norm are applied, and a different backend of ECAPA-TDNN Global~\cite{ECAPA}. In most of these settings, the proposed method consistently improves performance, yielding relative performance gains of 3.06\% to 4.53\%.

\subsection{Gender Classification}

We hypothesize that since INSIDE selects same-gender speaker pairs for identity interpolation, the resulting synthetic identities are likely to preserve the gender of the original pair. This suggests that INSIDE may also be beneficial for gender classification tasks. To evaluate this hypothesis, we conducted experiments on several datasets, and the results are shown in Table~\ref{tab_gender}. We observe that the proposed INSIDE-NN consistently outperforms the baseline without INSIDE-based data expansion in most cases, achieving an average improvement of 12.09\%. Furthermore, the ID-Exp variant, which introduces a larger number of synthetic identities, leads to even better performance with an average gain of 13.44\%.

\begin{figure}[h]
% \vspace{-0.1 in}
\centering
\includegraphics[width=1.01\columnwidth]{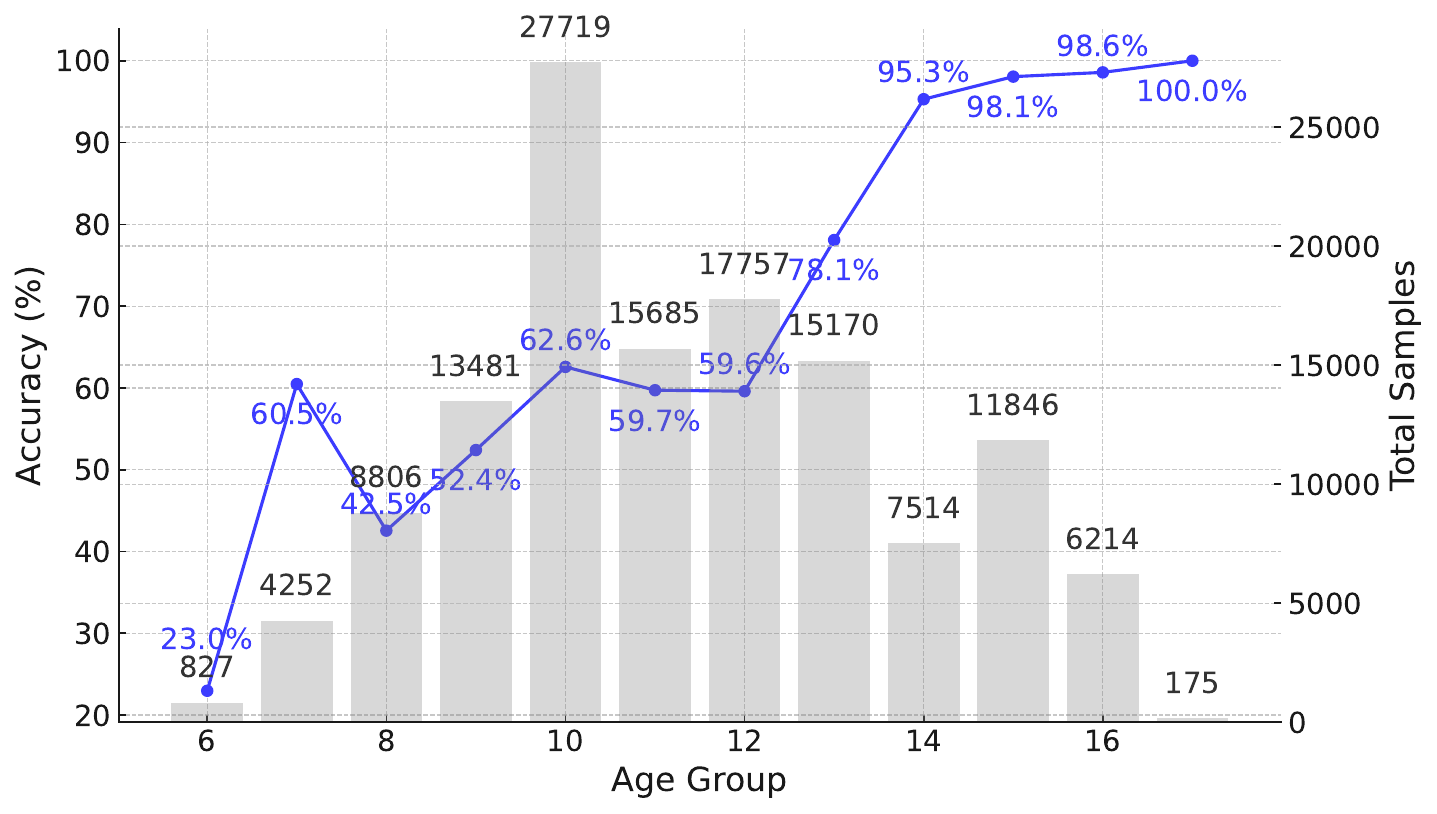}
\vspace{-0.3 in}
\caption{Accuracy across different age groups (blue line, left Y-axis) and corresponding number of samples in each group (gray bars, right Y-axis).}
\label{fig_ice_gender}
\vspace{-0.05 in}
\end{figure}

A deeper analysis reveals that the Private and Samrómur Children datasets exhibit higher classification error rates. One possible reason for this is language mismatch~\cite{10832142}, as both datasets differ from the English-based training data. Notably, Samrómur Children dataset shows high error rates. To further investigate this, we analyze the results by age group, as shown in Fig.~\ref{fig_ice_gender}. The findings indicate that gender classification is significantly more challenging for younger children (ages 6–13), with error rates increasing as age decreases. In contrast, for children aged 15 and above, the error rate drops below 2\%.

% \subsection{Towards a privacy-friendly Dataset}

\section{Limitations and Future Work}
Although the proposed INSIDE framework improves performance in both speaker verification and gender classification tasks, it still has several limitations.

First, the speaker encoder used in most current TTS systems is relatively lightweight and generally less accurate than those used in state-of-the-art speaker verification models based on speech foundation models~\cite{10096626}. This may limit the effectiveness of speaker embedding-based data expansion, particularly when applied to strong speaker verification baselines. For example, in our experiments with a highly competitive baseline, the relative performance gain is limited to 5.24\%. Future work could explore TTS models with more powerful speaker encoders to improve the quality of identity interpolation.

\begin{figure}[ht]
% \vspace{0.07 in}
\centering
\includegraphics[width=1\columnwidth]{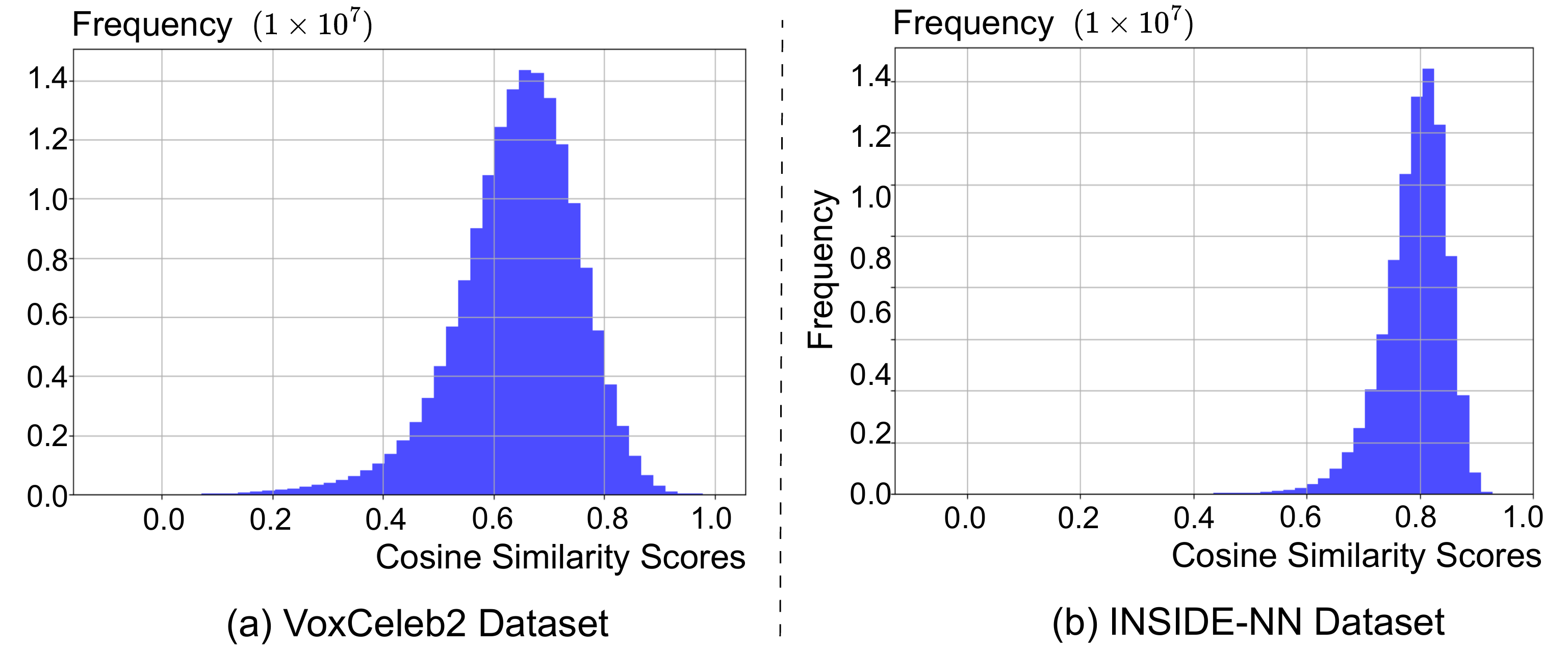}
\vspace{-0.25 in}
\caption{Distribution of cosine similarity scores between different utterances of the same identity in the (a) VoxCeleb2 dataset, and (b) INSIDE-NN dataset.}
\label{fig_distribution}
\vspace{-0.05 in}
\end{figure}

Second, as shown in Fig.~\ref{fig_distribution}, we observe that cosine similarity scores between different utterances from the same synthetic identity are significantly higher than those of real identities. This indicates that synthetic identities exhibit lower intra-class uncertainty compared to real speakers. Such a discrepancy in data distribution may pose challenges for speaker verification models, particularly in learning to model speaker uncertainty, and could negatively impact robustness~\cite{9463712}. Future work may explore generating synthetic data that better mimics the statistical properties of real speaker distributions, including intra-class variability, to reduce this mismatch.

\section{Conclusions}
In this work, we proposed INSIDE (Interpolating Speaker Identities in Embedding Space), a novel data expansion framework that generates new speaker identities via spherical interpolation between real speaker embeddings. By conditioning a TTS model on these interpolated embeddings, we synthesize realistic speech data, enhancing identity diversity without requiring additional data collection. We further introduce an optimized pair selection strategy and demonstrate INSIDE’s compatibility with different training configurations. Experimental results show that INSIDE consistently improves performance on both speaker verification and gender classification tasks, achieving up to 5.24\% and 13.44\% average relative gains, respectively.
Moreover, INSIDE is compatible with existing data augmentation methods and can further enhance performance when combined with them.

\section*{Acknowledgment}
{\fontsize{8.8pt}{10pt}\selectfont{Ke Zhang and Haizhou Li are supported in part by the Shenzhen Science and Technology Program (Shenzhen Key Laboratory, Grant No. ZDSYS20230626091302006), the Shenzhen Science and Technology Research Fund (Fundamental Research Key Project, Grant No. JCYJ20220818103001002), and the Program for Guangdong Introducing Innovative and Entrepreneurial Teams (Grant No. 2023ZT10X044).}

% \balance
\printbibliography
\end{document}